\documentclass[preprint]{iacrtrans}
\usepackage[utf8]{inputenc}

\usepackage{xcolor}
\usepackage{extarrows}
\usepackage{braket}
\usepackage{soul}
\let\underline\ULunderline
\usepackage{graphicx}
\usepackage{tikz}
\usepackage{tabularx}
\usepackage{stmaryrd}
\usepackage{comment}
\usepackage{dsfont}
\usepackage{amsthm,amsmath,amssymb,mathrsfs,amstext,amsfonts}
\usepackage[OT1]{fontenc}

\usepackage{tcolorbox}
\tcbset{
  colframe=black!80,
  colback=gray!10,
  boxrule=0.3mm,
  arc=2mm,
  center title,
  boxsep=1pt,       
  left=1mm, right=2mm, top=1mm, bottom=1mm  
}
\usepackage{float}
\usepackage{caption}

\usepackage{multirow}
\usepackage{booktabs}
\usepackage{mdframed}
\usepackage{dirtytalk}
\usepackage{algorithm}
\usepackage{algorithmic}
\usepackage{graphicx}
\usepackage{float}
\usepackage{enumitem}
\usepackage{multicol}
\usepackage{url}

\usepackage[normalem]{ulem}
\definecolor{LimeGreen}{rgb}{0.2, 0.8, 0.2}
\usepackage{todonotes}

\usepackage[psdextra]{hyperref}
\usepackage{cleveref}

\newcommand{\splitatcommas}[1]{%
    \begingroup
    \begingroup\lccode`~=`, \lowercase{\endgroup
        \edef~{\mathchar\the\mathcode`, \penalty0
            \noexpand\hspace{0pt plus 1em}}%
        }\mathcode`,="8000 #1%
    \endgroup
}

\newcounter{protocol}

\begin{document}
\title[Improving Quantum Oblivious Transfer using Zero-Knowledge Proofs]{Improving Quantum Oblivious Transfer using Zero-Knowledge Proofs}

 \author{Alex B. Grilo$^1$ \and Lucas Hanouz$^{1,2,*}$
 \and Anne Marin$^2$ }
 \institute{
 Sorbonne Université, CNRS, LIP6, Paris, France
 \and
 VeriQloud, Paris, France
 }
\maketitle

\vspace{-1.2cm}
\begin{center}
    \small{$^{*}$ Corresponding author: lucas.hanouz@lip6.fr}
\end{center}


\begin{abstract}
While quantum protocols for oblivious transfer can be built from weaker assumptions than classical ones, the current protocols remain highly impractical due to stringent security bounds that demand a large amount of quantum communication.

In this work, we propose an enhancement to quantum oblivious transfer protocols by leveraging zero-knowledge proofs (ZKPs). At a high level, we modify the protocol such that the receiver proves that they are following honestly some crucial steps of the protocols, without revealing extra information to the sender. This yields a tighter security bound against a malicious receiver. 

For practical scenarios with a fixed security target, this approach reduces the amount of required quantum communication by one to two orders of magnitude, depending on the specific parameters. Furthermore, the benefits of our approach become particularly pronounced at higher qubit error rates, and our proposal even extends the maximum tolerable error rate of the quantum channel from approximately $3.5\%$ to about $9\%$.

The ZKP, however, introduces a substantial computational overhead, estimated to significantly exceed the acquisition time of the protocol, which is usually the primary bottleneck. To address this limitation, we also introduce a variant that serves as a tunable trade-off between the protocol with and without our modification. This enables precise adjustment of the computational overhead while still retaining significant, albeit moderately reduced, efficiency gains.
\end{abstract}

\section{Introduction}

Oblivious Transfer (OT) is a fundamental cryptographic primitive that enables secure multiparty computation~\cite{OT-SMPC}. It allows a receiver to securely obtain one of two messages provided by a sender, ensuring that the receiver learns nothing about the other message, while the sender remains oblivious to the receiver's choice. Although oblivious transfer is known to require strong computational assumptions classically, it has been shown~\cite{OTfromOWF1,OTfromOWF2} that using quantum resources, the assumption of One-Way Functions (OWF) is sufficient to obtain composable security in the plain model, a highly desirable security property that allows OT to remain secure if used inside a larger cryptographic construction such as secure multiparty computation. 

However, the original protocols were impractical, and, despite the recent efforts to design more efficient protocols~\cite{Y24}, the estimated performance remains barely practical even in very optimistic scenarios. Other proposals relax the security constraints to improve efficiency, sacrificing composability and relying on the random oracle model~\cite{VQ_OT_Tokens}. Even so, reported execution rates do not exceed a few oblivious transfers per hour, classifying these implementations as proofs of concept rather than practical solutions.\\

In this work, we propose an enhancement to Quantum Oblivious Transfer (QOT) protocols leveraging zero-knowledge proofs. ZKPs constitute a foundational cryptographic primitive that enables a prover to convince a verifier of the validity of a statement without revealing any information beyond its intrinsic truth.

In known QOT protocols, the uncertainty principle is exploited to restrict the receiver to accessing approximately half of the sender's secrets. Within these schemes, the receiver partitions these secrets into two sets, which the sender subsequently uses to mask the two messages. While an honest receiver concentrates all their acquired knowledge into one of the two sets, security proofs rely on a bound of the receiver's entropy on the lesser-known set and must account for the worst-case scenario wherein a malicious receiver would split their knowledge evenly across both sets.
Our proposal consists of a protocol modification where the receiver provides a ZKP certifying an honest partition, effectively preventing this even split.\\

The main consequence is an improvement of the security bounds of QOT protocols against a malicious receiver.
We present the improved security bound in an error-tolerant scenario and assuming external leaks, but our result extends to other scenarios as long as the protocols follow the same structure.

In practice, when the target bit security is fixed, this improvement translates into a drastic reduction in the number of qubits that need to be transmitted from the sender to the receiver. This directly decreases the state distribution (or acquisition) time, as well as the original post-processing time and memory usage. While the precise magnitude of this improvement depends on various factors, realistic parameterizations yield a reduction in required quantum communication by a factor of $18$ at $1\%$ qubit error rate, and $199$ at $3\%$.
Furthermore, our approach raises the maximum tolerable error rate of the quantum transmission from approximately $3.5\%$ to around $9\%$. Although a $3.5\%$ threshold is well within the capabilities of modern quantum communication hardware, this increased tolerance could make QOT viable via lower-end instruments or over greater distances.\\

These improvements, however, come at the cost of substantial computational overhead due to the use of a ZKP over a statement involving heavy computations. Our estimations indicate that, in the absence of dedicated hardware acceleration for classical computations, this overhead can easily represent a computational time several orders of magnitude greater than the improved acquisition time, and still one order of magnitude greater than the acquisition time of the original protocol, without our proposed modification.

To address this issue, we introduce an efficient alternative to our proposal, which serves as a tunable intermediate between our initial proposal and the original protocol. This alternative reduces the computational workload of the ZKP by relaxing the proven statement. 
This variant allows for precise adjustment of the computational overhead, albeit at the expense of partially sacrificing the aforementioned gains. 
We provide a comparison of the performances of this alternative with the improvements and computational overhead of the initial proposal, and of the original protocol. Notably, we show that under realistic parameters, the computational overhead can be reduced to match the acquisition time while still yielding significant improvements over the baseline protocol. 

Finally, we emphasize that should future advances in the computational efficiency of ZKPs occur, they will inherently benefit our approach by directly alleviating the overhead introduced by our modification.\\

\vspace{3mm}
The remainder of this work is structured as follows. 
\Cref{sec:prelim} presents preliminaries. \Cref{sec:OT_background} presents the general structure of QOT protocols that our work improves. \Cref{sec:ZK} presents our proposal and its impact on the final security of QOT. \Cref{sec:benchmark} provides an analysis of the gains on the quantum communication of our proposal. In \Cref{sec:est_and_mitig_overhead}, we estimate the computational overhead implied by our proposal and introduce our alternative proposal for mitigating this overhead.
Finally, we explore future directions in \Cref{sec:future_directions}

\paragraph{Acknowledgements.}
ABG is supported by
the European Union’s Horizon Europe Framework Programme under the Marie Skłodowska
Curie Grant No. 101072637, Project Quantum-Safe Internet (QSI) and by ANR JCJC TCS-NISQ ANR-22-
CE47-0004.

\section{Preliminaries} \label{sec:prelim}

We assume basic knowledge about the theory of quantum communication and computing.

\subsection{Useful lemmas}
\subsubsection{ Leftover Hash Lemma }
We will make use of the following Leftover Hash Lemma with quantum side information, introduced in a more general form in \cite{LeftoverHashLemma2}.
\begin{lemma} \label{lem:leftover_hashing}
    Let $X$ be a random variable, let $E$ be a quantum system, and let $F$ be a family of two-universal hash functions from $X$ to $\{0, 1\}^\ell$. Then, on average over the choices of $f$ from $F$, the output $Z := f(X)$ is $\Delta$-close to uniform conditioned on $E$, where
    \begin{equation}
        \Delta = \frac12\sqrt{2^{\ell-H_{min}(X|E)}}
    \end{equation}
    with $H_{min}(X|E)$ denoting the min-entropy of $X$ conditioned on $E$, i.e. $H_{min}(X|E) = -\log(p_{\mathrm{guess}}(X|E))$ where $p_{\mathrm{guess}}(X|E)$ is the maximum  probability of correctly guessing the value of $X$ using the optimal strategy with access to $E$.
\end{lemma}

\subsubsection{ Hoeffding's bound }
On several occasions, given a set $T$, we will consider random subsets of $T$, that are sampled by independently selecting each element with a fixed probability. In these contexts, we will need to bound the probability that the sizes of such subsets deviate from their expected values by more than a certain margin $\delta$. To this end, we will rely on Hoeffding's inequalities~\cite{Hoeffding}, which, when applied to $n$ independent Bernoulli random variables, provide the following upper bounds on the probability that the sum $S$ of these variables deviates from its expected value $E[S]$ by more than a certain margin $\delta$:
\begin{equation}
    P\left[\left|S-E[S]\right|\ge \delta\right] \le 2e^{-2\frac{\delta^2} n}
\end{equation}
or equivalently
\begin{equation}
    P\left[\left|S-E[S]\right|\ge \delta n\right] \le 2e^{-2\delta^2n}.
\end{equation}

Applied to a subset $T_1$ of a set $T$ of size $n$, considering the $n$ random variables $(X_i)_{i\in[n]}$ taking the value $1$ if the $i$-th element of $T$ is in $T_1$, and $0$ otherwise, these inequalities give
\begin{equation}
    P\bigg[\Big||T_1|-E[|T_1|]\Big|\ge \delta n\bigg] \le 2e^{-2\delta^2n}.
\end{equation}
We will refer to general Hoeffding's inequalities when invoking this result.

\subsection{Zero-knowledge proof systems}\label{sec:prelim_ZK}

Zero-Knowledge Proofs (ZKPs) are cryptographic constructions that allow a prover to prove a statement to a verifier in such a way that no information beyond the validity of the statement is leaked.

Let $R$ be an NP relation consisting of pairs $(x, w)$, where $x$ is the public statement and $w$ is the corresponding private witness. We define the associated language as $L_R = \{x \mid \exists w : (x, w) \in R\}$. An interactive proof system for $L_R$ is a protocol executed between two probabilistic polynomial-time (PPT) Turing machines: a Prover $P$ and a Verifier $V$. 

At the conclusion of their interaction over a common input $x$, denoted as the transcript $\langle P(w), V \rangle(x)$, the Verifier outputs a decision bit $b \in \{0, 1\}$, where $1$ denotes acceptance and $0$ denotes rejection.

A zero-knowledge proof system must fundamentally satisfy the three following properties:
\begin{itemize}
    \item \textbf{Completeness:} If $(x, w) \in R$, an honest Prover $P$ interacting with an honest Verifier $V$ will convince $V$ to accept with overwhelming probability.
    \item \textbf{Soundness:} If $x \notin L_R$, no malicious prover $P^*$ can convince the honest Verifier $V$ to accept, except with a negligible probability $\epsilon_{S}$, termed the soundness error.
    \item \textbf{Post-quantum zero-knowledge:} 
    
    A proof system is \emph{post-quantum zero-knowledge} if, for any efficient (quantum) malicious verifier $V^*$, there exists an efficient algorithm $S$ (called the \emph{simulator}) that, without access to the witness, can reproduce the final internal state of $V^*$ at the end of a real interaction with an honest prover holding a valid witness.\\
    More precisely, a system is post-quantum zero-knowledge with distinguishing advantage $\epsilon_{ZK}$ if, for any valid statement and any auxiliary input provided to $V^*$ and $S$ prior to the protocol, any efficient distinguisher has advantage at most $\epsilon_{ZK}$ in distinguishing the state output by $S$ from the final internal state of $V^*$ after interacting with the prover.
    
\end{itemize}

It is important to note the following:
Classically, zero-knowledge proofs have long been known to be achievable from one-way functions~\cite{classical_ZKPoK_from_OWFs}. 
Post-quantum zero-knowledge proofs are, however, famously harder to obtain.
Watrous~\cite{Watrous_rewinding} proved that quantum-secure zero-knowledge proofs can be obtained from one-way functions by using constructions that sequentially realize multiple interactive passes which individually feature non-negligible cheating probability.

\section{Background on quantum protocols for OT} \label{sec:OT_background}

\subsection{The QOT protocol}\label{sec:protocol}

The now standard Quantum Oblivious Transfer protocol, initially proposed by Crepeau and Killian~\cite{CK88IEEE} and Bennett, Brassard, Crépeau and Skubiszewska~\cite{QOT}, consists of the following steps where Alice is the sender and Bob is the receiver.

\noindent \textbf{Parameters: } $k, n \in \mathbb{N}$ two integers with $k<n$. $l \in \mathbb{N}$ the bit-size of the input messages.
\noindent \textbf{Inputs: } \textul{Alice:} two messages $m_0, m_1 \in \{0,1\}^\ell$ ; \textul{Bob:} a choice bit $b\in \{0,1\}$\\
\textbf{Outputs: } \textul{Alice:} $\emptyset$ ; \textul{Bob:} $m_b$
\begin{enumerate}
    \item \textbf{State distribution:} Alice samples random bit strings $(x^A_i)_{i\in[n]}$ and $(\theta^A_i)_{i\in[n]}$ and sends the state $\bigotimes_{i\in[n]}H^{\theta^A_i}\ket{x^A_i}$ to Bob, where $H$ denotes the Hadamard gate. 
    
    Bob samples random bases $(\theta^B_i)_{i\in[n]}\in\{0, 1\}^n$ and measures each received qubit in the corresponding basis (computational basis if $\theta^B_i=0$ or Hadamard basis otherwise), denoting the measurement outcome $(x_i^B)_{i\in[n]}$. 
    
    \item \textbf{Commitments:} Bob commits to each of the measurement outcomes (by associating each with its indices) using a commitment scheme. We denote by $(c_i^\theta)_{i\in[n]}$ and $(c_i^x)_{i\in[n]}$ the commitments of the basis and the secret measurement outcomes, respectively.
    
    \item \textbf{Parameter testing and basis revelation:} Alice samples and announces a random subset of size $k$, denoted $T\subset[n]$. Bob opens his commitments corresponding to the indices in $T$ and Alice computes the error rate (relative Hamming distance) between the opened secrets and her corresponding ones, restricted to the indices for which the bases match ($\theta^A_i = \theta^B_i$). 
    If the error rate exceeds a threshold $\alpha$, they abort. 
    
    Alice announces her bases  $(\theta^A_i)$ for the remaining indices.
    
    \item \textbf{Set partitioning:} Bob partitions the remaining indices into two disjoint sets $I_0$ and $I_1$ such that the set $I_b$ contains all indices for which his measurement basis matches Alice's basis ($\theta^B_i = \theta^A_i$) and $I_{\bar b}$ contains the other indices. 
    He announces $I_0$ and $I_1$ to Alice.

    \item \textbf{Post-processing (Alice):} From her secret strings $x^A_{I_0}$ and $x^A_{I_1}$, Alice computes:
    \begin{itemize}
        \item error-correcting syndromes $s_0$ and $s_1$ that she sends to Bob,
        \item the $\ell$-bit keys $\tilde x_0$ and $\tilde x_1$, by privacy amplification, i.e. applying a 2-universal hash function to $x^A_{I_0}$ and $x^A_{I_1}$.
    \end{itemize}
    Alice then encrypts her two messages into ciphertexts $C_0 = m_0 \oplus \tilde  x_0$ and $C_1 = m_1 \oplus \tilde x_1$ which she sends to Bob. 

    \item \textbf{Post-processing (Bob):} 
    Bob then uses the syndrome $s_b$ to correct his string $x^B_{I_b}$ and compute the key $\tilde x_b$. He then obtains his final output as $m_b = C_b\oplus \tilde x_b$.
\end{enumerate}

Quantum Oblivious Transfer protocols following the presented structure have been proven secure in several different settings~\cite{Yao_QOT_proof95,BF12,OTfromOWF1,OTfromOWF2,Y24}.
 
The primary distinction among various instantiations of QOT protocols lies in
the choice of the underlying bit commitment scheme. While most frameworks employ
\textit{statistically binding} commitment schemes~\cite{Yao_QOT_proof95,BF12},
others rely on more elaborate variants~\cite{OTfromOWF1,OTfromOWF2,Y24,DFL09} to
achieve \textit{composable} security, a highly desirable property for OT which
ensures that the protocol remains secure when deployed as a building block
within broader cryptographic constructions, such as MPC
protocols~\cite{OT-SMPC,IPS08}. However, this incurs a substantial cost, as these specific commitments require either strong computational assumptions or, if instantiated under weaker assumptions such as one-way functions, the use of quantum protocols with a significant overhead in the transmission of quantum states.

In this work, we focus on statistically binding commitments for simplicity, but our result extends to the composable frameworks of \cite{OTfromOWF1,OTfromOWF2,Y24,DFL09}.

\subsection{Security}

We note that the modifications that we introduce in the protocol will
barely change the security against a malicious Alice, so we focus here on the security against a malicious Bob.

Throughout this work, we use the following security definition for oblivious transfer against a malicious Bob.
An OT protocol is $\epsilon$-secure against a malicious
Bob if there exists a bit $b\in\{0, 1\}$ such that Alice's key $\tilde x_b$ is $\epsilon$-close to uniformly random, and independent of Bob’s view.

\begin{definition}[Security of OT against a malicious Bob]\label{def:OT_sec}
    An OT protocol is said to be $\epsilon$-secure against a malicious Bob if
    there exists a bit $b\in\{0, 1\}$ such that the state $\rho_{\tilde x_{\overline{b}}B}$ consisting in the view $\rho_{B}$ of a malicious Bob at the end of the protocol, along with Alice's key $\tilde x_{\overline{b}}$, satisfies:
    \begin{equation}
      \Delta(\rho_{\tilde x_{\overline{b}}B}, \frac 1{2^\ell}\mathbb{I}\otimes\rho_{B}) \le \epsilon,
    \end{equation}
    where $\Delta$ denotes the trace distance.
\end{definition}

In the remainder of the paper, we will refer to $\epsilon$ either as the distinguishing advantage of a malicious Bob or as the security bound against a malicious Bob.

Our work is based on the proof of Bouman and Fehr~\cite{BF12} which is conducted in the noiseless setting. 

They bound the distinguishing advantage of a malicious Bob for QOT protocols following the structure of \Cref{sec:protocol} in the noiseless setting for any $\gamma, \delta > 0$ by  
\begin{equation}\label{eq:BF12sec}
    \epsilon_{\cite{BF12}} 
    \leq \frac12 \times 2^{ -\frac12\Big( \Big(\frac14-\frac\gamma2-h(\delta)\Big)(n-k) -\ell \Big) } + \sqrt6 e^{-\frac{\delta^2k}{100}} + 2e^{-2\gamma^2(n-k)}
\end{equation}
where $n, k, \ell$ are the parameters of the protocol (see \Cref{sec:protocol}). \\

By adapting the security proof of~\cite{BF12} to an error-tolerant setting, and considering their proposed $\chi$-relaxed extractable commitment scheme as well as a fixed number of check qubits $k=\frac n2$, the work of \cite{Y24} bounds the distinguishing advantage of a malicious Bob for their protocol by:

\begin{equation} \label{eq:Y24sec}
    \epsilon_{\cite{Y24}} \leq
    \frac{1}{2} \cdot 2^{-\frac{1}{2}\Big[\left[\frac{1}{4}-\gamma-\vartheta-\frac12h(\delta+\alpha+\chi)(1-2\vartheta)\right](n-k)-\ell-q\Big]} + \sqrt6 e^{-\frac{\delta^2k}{100}} + 2e^{-2\gamma^2(n-k)}
\end{equation}
where:
\begin{itemize}
    \item $\alpha$ is the qubit error rate,
    \item $q$ is the size of the syndromes used for the error-correcting step,
    \item $\vartheta$ is the proportion of leaked bits (for instance due to multi-photon events).
\end{itemize}

\section{Improving the security bound against malicious Bob} \label{sec:ZK}

Our proposed modification is to add the partition check step described below, after the set partitioning step and before the post-processing step.
This additional step is parametrized by a margin $\mu$ which represents the maximum tolerable (relative) deviation for the size of the sets $I_0$ and $I_1$, compared to their expected value $\frac{n-k}2$.

\begin{enumerate}[label=4\alph*, start=2]
    \item \textbf{Partition check:} \\
    \vspace{3mm}\hspace{4mm}\textit{Parameters: $\mu \in [0, 1]$}\\
    If $|I_0| < \frac{n-k}{2}(1-\mu)$ or $|I_1| < \frac{n-k}{2}(1-\mu)$, the protocol aborts.
    
    Bob computes and sends to Alice a zero-knowledge proof $\pi$ that 
    \begin{equation} \label{ZK-statement}
    \exists b,(o_i)_{i\in I_{\bar b}} \forall i \in I_{\bar b}, \textbf{Open(}c_i^\theta,o_i\textbf{)}=\overline{\theta_i^A}
    \end{equation}
    where $\textbf{Open(}c_i^\theta,o_i\textbf{)}$ is the opening of the commitment $c_i^\theta$ with the opening key $o_i$.
    
    Alice verifies the proof, aborting if it fails.
\end{enumerate}

In the remainder of this section, we show that this additional step improves the security of Quantum Oblivious Transfer against malicious Bob, while not disturbing correctness and security against malicious Alice significantly.

\subsection{Correctness}
Notice that our modification introduces two additional abort conditions to the Oblivious Transfer protocols. We argue in the following \Cref{thm:correctness} that the probability for the new protocol to abort remains negligible if the players are honest.
\begin{theorem}\label{thm:correctness}
    Let $\epsilon_{cor, OT}$ be the correctness error of a quantum oblivious transfer protocol following the structure described in \Cref{sec:protocol}.
    The resulting protocol with the added partition check step has correctness error 
    $$\epsilon_{cor, OT}+\epsilon_{cor, ZK}+\epsilon^\mu$$
    where 
    \begin{itemize}
        \item $\epsilon_{cor, ZK}$ is the correctness error of the zero-knowledge proof system,
        \item $\epsilon^\mu = 2e^{-\frac{n-k}{2}\mu^2}$.
    \end{itemize}
\end{theorem}

\begin{proof}
    The partition check step adds two new aborting conditions.
    First, Alice may reject the honest zero-knowledge proof. By definition, this happens with probability $\epsilon_{cor, ZK}$, the correctness error of the zero-knowledge proof system.
    Secondly, the protocol aborts if $|I_0| < \frac{n-k}{2}(1-\mu)$ or if $|I_1| < \frac{n-k}{2}(1-\mu)$. Since, in an honest scenario, $I_0$ and $I_1$ form a uniformly random splitting of $n-k$ bits, $|I_0|+|I_1|=n-k$, $\mathds{E}[|I_0|]=\mathds{E}[|I_1|] = \frac{n-k}{2}$ and $P\left[|I_0| < \frac{n-k}{2}(1-\mu)\lor|I_1| < \frac{n-k}{2}(1-\mu)\right] = P\left[\left||I_0| - \frac{n-k}{2}\right| > \mu\frac{n-k}2\right]$. We conclude the proof by applying Hoeffding's inequality, which ensures that
    \begin{equation*}
        P\left[\left||I_0| - \frac{n-k}{2}\right| \ge \mu\frac{n-k}2\right]
        \le 
        2e^{-\frac{n-k}{2}\mu^2}. \qedhere
    \end{equation*}
\end{proof}

\subsection{Security against malicious Alice}

We now show how the addition of the partition check step changes the security against a malicious sender. Here, we adopt the following simulation-based security definition.

\begin{definition}[Security of OT against a malicious sender] \label{def:sec_OT_malicious_Alice}
    An OT protocol is $\epsilon$-secure against a malicious sender if there
    exists a simulator $S$ such that any polynomially bounded distinguisher has an advantage at most $\epsilon$ in distinguishing the sender's view of the real protocol from the one of the joint system of $S$ and an ideal OT functionality $\mathcal{F}_{OT}$.
\end{definition}

\begin{theorem} \label{thm:malicious_alice}
    Consider a QOT protocol following the structure described in \Cref{sec:protocol} that is $\epsilon^B_{QOT}$-secure against a malicious sender under \Cref{def:sec_OT_malicious_Alice}. 
    Then, the same protocol with the additional partition check step is $ \epsilon_B$-secure against a malicious sender, with
    \begin{equation} \label{eq:malicious_alice_sec}
        \epsilon_B \leq \epsilon_{zk} + \epsilon^B_{QOT},
    \end{equation}
    where $\epsilon_{zk}$ is the zero-knowledge error of the employed zero-knowledge proof system.
\end{theorem}

\begin{proof}

We proceed via a sequence of hybrid games to bound the advantage of a polynomially bounded distinguisher taking the role of a malicious sender.

Let $\mathcal{S}_{QOT}$ be a simulator such that a given class of distinguishers has an advantage of at most $\epsilon^B_{QOT}$, in distinguishing the sender's view of the real protocol from the one of an ideal OT functionality $\mathcal{F}_{OT}$ and the simulator $\mathcal{S}_{QOT}$.
Consider a distinguisher of this class.\\

\noindent \textbf{Hybrid 0:} This represents the execution of the protocol with the added partition check step. An honest receiver computes the zero-knowledge proof $\pi$ using their actual witness $(b, (o_i)_{i\in I_{\bar b}})$. Let $\text{View}_{\mathcal{A}^*}^0$ denote the view of the distinguisher in this execution.

\noindent  \textbf{Hybrid 1:} We modify the partition check step by substituting the zero-knowledge proof with a simulated one. Specifically, we invoke the simulator $\mathcal{S}_{ZK}$ associated with the zero-knowledge proof system. Given only the public statement from \Cref{ZK-statement}, $\mathcal{S}_{ZK}$ generates a simulated proof $\tilde{\pi}$ (without access to Bob's witness). Let $\text{View}_{\mathcal{A}^*}^1$ denote the view of the distinguisher in this modified execution.

\noindent \textbf{Hybrid 2:} This represents the execution of the ideal OT functionality $\mathcal{F}_{OT}$ and the simulator $\mathcal{S}_{QOT}$, on top of which the simulator $\mathcal{S}_{ZK}$ of the zero-knowledge proof system is used to generate a simulated proof $\tilde{\pi}$.\\

By the zero-knowledge property of the proof system, the advantage of the distinguisher in distinguishing the simulated proof $\tilde{\pi}$ from the real proof $\pi$ is at most $\epsilon_{zk}$. Consequently, so is her advantage in distinguishing the views $\text{View}_{\mathcal{A}^*}^0$ and $\text{View}_{\mathcal{A}^*}^1$ of Hybrids 0 and 1, respectively.

The simulated proof $\tilde{\pi}$ is generated independently of the remainder of the protocol, and the added partition check step only adds this proof to the view of the distinguisher compared to the original protocol. As a consequence, the advantage of the distinguisher in distinguishing the views $\text{View}_{\mathcal{A}^*}^1$ and $\text{View}_{\mathcal{A}^*}^2$ of Hybrids 1 and 2 is exactly her advantage in distinguishing the original QOT protocol from the ideal OT functionality $\mathcal{F}_{OT}$ and the simulator $\mathcal{S}_{QOT}$. By assumption, this advantage is at most $\epsilon^B_{QOT}$.

Finally, the advantage of the distinguisher in distinguishing Hybrids 0 and 2, i.e. the security bound of a QOT protocol with the added partition check step against a malicious sender, is bounded by $\epsilon_B \leq \epsilon_{zk} + \epsilon^B_{QOT}$.
\end{proof}

\subsection{Security against malicious Bob} \label{sec:FullZKsecurity}

From the security proof of QOT of \cite{BF12}, we derive \Cref{thm:malicious_bob_Y24_improved}, which gives an improved security bound against a malicious Bob for QOT protocols with the additional partition check step. 
As detailing the formal proof here would entail reproducing most of the technical arguments developed in \cite{BF12}, we instead provide a proof sketch. This sketch outlines the intuition behind the original proof and explicitly highlights the deviations introduced by our partition check step.\\

We now state a theorem that follows from the results in \cite{BF12}\footnote{More concretely, \Cref{thm:BF12_Hmin_bound_imported} combines the results in Corollary 1, Theorem 3, and Example 4 from \cite{BF12}.}.

\begin{theorem}\label{thm:BF12_Hmin_bound_imported}
    
    Consider a quantum system $AE$ where $A$ is an $n$-qubit subsystem. Let:

    \begin{itemize}
        \item $(\hat \theta_i)_{i\in[n]}, (\hat x_i)_{i\in[n]} \in \{0,1\}^n$ be fixed.
        \item $k\in[n-1]$ and let $T$ be a random subset of size $k$ of $[n]$.
        \item $\theta=(\theta_i)_{i\in [n]}\in \{0,1\}^n$ be an $n$-bit string where the variables $\theta_T$ corresponding to indices in $T$ are uniformly random.
        \item $I$ be any subset of $[n]\backslash T$.
        \item $W$ and $X$ be the random variables obtained by measuring qubits $i\in T$ and $i\in I$, respectively, each in basis $H^{\theta_i} \{\ket0, \ket1\}$.
    \end{itemize}
    
    Suppose that 
    \begin{equation}
        \frac{|\{i\in T \text{ such that } \theta_i = \hat\theta_i \text{ and } W_i \neq \hat x_i\}|}{|\{i\in T \text{ such that } \theta_i = \hat\theta_i\}|} \le \alpha.
    \end{equation}

    Then, for any $\delta\in[0, \frac12-\alpha]$, except with probability $\epsilon^\delta = \sqrt6 e^{-\frac{\delta^2k}{100}}$, the min-entropy $H_{min}(X|E)$ on the variable $X$ given system $E$ satisfies
    \begin{equation}
        H_{min} (X|E) \ge d(\theta_{I}, \hat \theta_{I}) - h(\delta + \alpha)\times |I|
    \end{equation}
    where $d(\theta_{I}, \hat \theta_{I})$ denotes the Hamming distance between $(\theta_i)_{i\in I}$ and $(\hat \theta_i)_{i\in I}$.
\end{theorem}

We now show in \Cref{thm:malicious_bob_Y24_improved} the improvement on the security against malicious Bob in the presence of errors and leaks.

\begin{theorem} \label{thm:malicious_bob_Y24_improved}
    Consider a QOT protocol following the structure described in \Cref{sec:protocol} with statistically binding commitments and the additional partition check step parametrized by $\mu$.
    Then, for any $\delta \in [0, \frac12]$, the protocol is $\epsilon_A$-secure against a malicious Bob in the sense of \Cref{def:OT_sec} with
    \begin{equation} \label{eq:Y24sec_improved}
        \epsilon_A \leq
        \frac{1}{2} \cdot 2^{-\frac{1}{2}\Big[\frac{n-k}{2}(1-\mu-2\vartheta)(1 - h(\delta+\alpha))-\ell-q\Big]} + \sqrt6 e^{-\frac{\delta^2k}{100}}+\epsilon_S
    \end{equation}
    where:
    \begin{itemize}
        \item $\alpha$ is the qubit error rate,
        \item $q$ is the size of the syndromes used for the error-correcting step,
        \item $\vartheta$ is the proportion of Alice's secret bits which are leaked to Bob (for instance due to plausible multi-photon events),
        \item $\epsilon_S$ is the soundness error of the zero-knowledge proof.
    \end{itemize}

\end{theorem}

\begin{proof}[Proof sketch.]

We extend the security proof of QOT against malicious Bob from~\cite{BF12}, adapting the critical steps to accommodate our specific setting with noise and the additional partition check step.
The proof considers a purified version of the protocol, called $QOT^*$, where Alice sends halves of EPR pairs instead of BB84 states and chooses her basis (now for the measurement of the halves of EPR pairs that she kept) at the last moment in the parameter testing step. We write $\ket{\varphi_{AE}}$ the joint state held by Alice and malicious Bob. Without loss of generality, we may assume that $\ket{\varphi_{AE}}$ is indeed pure.
By statistical binding of the commitment scheme, Bob's commitment bases $\theta^B$ are uniquely determined.

By soundness of the zero-knowledge proof, if Alice accepts the proof, the proven Statement~\ref{ZK-statement} must be true except with probability $\epsilon_{S}$. For the remainder of the proof, we assume that there exists a valid witness $(b, (o_i)_{i\in I_{\bar b}})$ and we book-keep the probability.

Statement~\ref{ZK-statement} implies that
\begin{equation} \label{eq:wt_theta_I1-c__leaks_improved}
    d(\theta^A_{\bar b}, \theta^B_{\bar b}) = |I_{\bar b}|.
\end{equation}

We assume external leakage of a proportion $\vartheta$ of Alice's secrets among the $n-k$ indices of $I_0\cup I_1$\footnote{The leakage is modeled as the exposure of $\vartheta (n-k)$ explicit bits of Alice's secret, which are considered to be perfectly known by the distinguisher. This stands in contrast to a model where the distinguisher would acquire $\vartheta (n-k)$ bits of generalized information on the secret. This discrete leakage notably aims at capturing multi-photon events in practical implementations.}.
Considering that malicious Bob is taking maximum advantage of these $\vartheta(n-k)$ leaks by putting all of their corresponding indices in $I_{\bar b}$, we will focus on the $|I_{\bar b}|-\vartheta(n-k)$ bits of $I_{\bar b}$ that didn't leak. We call $S$ this subset.

\Cref{eq:wt_theta_I1-c__leaks_improved} implies that
\begin{equation}
    d(\theta^A_{S}, \theta^B_{S}) = |S|.
\end{equation}

Let $X_{S}$ be the random variable obtained by measuring the qubits of indices in $S$ in basis $H^{\theta^A_{S}} \{\ket0, \ket1\}^{\otimes|S|}$.
Given the structure of the parameter testing step in the noise-tolerant setting, \Cref{thm:BF12_Hmin_bound_imported} can be applied to $X_{S}$, yielding
\begin{equation}
    H_{min} (X_S|E) \ge  d(\theta^A_{S}, \theta^B_{S}) - h(\delta+\alpha) \times |S|.
\end{equation}
As a consequence, since $S$ is a subset of $I_{\bar b}$,
\begin{align} \label{eq:Hmin_new_Y24_nochi}
    H_{min} (X_{\bar b}|E) 
    &\ge H_{min} (X_S|E)\\
    &\ge |S| \times ( 1 - h(\delta+\alpha))\\
    &\ge (|I_{\bar b}|-\vartheta(n-k)) \times ( 1 - h(\delta+\alpha)).
\end{align}

Remembering the soundness error probability, the proof is concluded by applying the leftover hashing lemma (\Cref{lem:leftover_hashing}), considering the leakage $q$ from the syndrome of the error correction and the bound $|I_{\bar b}| \ge \frac{n-k}{2}(1-\mu)$.

\end{proof}

\Cref{thm:malicious_bob_Y24_improved} equivalently applies to a setting without noise or leaks by setting the corresponding parameters ($\alpha$ or $\vartheta$ respectively) to $0$, and by considering that in the absence of noise, the part of the post-processing concerning the error correction syndromes $s_0$ and $s_1$ is skipped.
\begin{corollary}
    In a perfect setting without losses nor errors, consider a QOT protocol following the structure described in \Cref{sec:protocol} with statistically binding commitments and the additional partition check step parametrized by $\mu$.
    Then, for any $\delta \in [0, \frac12]$, the protocol is $\epsilon_{A, \text{noiseless}}$-secure against a malicious Bob in the sense of \Cref{def:OT_sec} with
    \begin{equation}
        \epsilon_{A, \text{noiseless}} \leq
        \frac{1}{2} \cdot 2^{-\frac{1}{2}\Big[\frac{n-k}{2}(1-\mu)(1 - h(\delta))-\ell\Big]} + \sqrt6 e^{-\frac{\delta^2k}{100}}+\epsilon_S
    \end{equation}
    where:
    \begin{itemize}
        \item $q$ is the size of the syndromes used for the error-correcting step,
        \item $\epsilon_S$ is the soundness error of the zero-knowledge proof
    \end{itemize}
\end{corollary}

\section{Transmission gains}\label{sec:benchmark}

In this section, we provide estimations of the transmission gains implied by our proposal. By considering fixed parameters and final security, we study how the resources requirements  vary with the introduction of our partition check step.

A quick comparison between \Cref{eq:Y24sec_improved,eq:Y24sec} already suggests that the same security can be achieved with about half as many rounds of quantum communication in a low-error regime. By carefully minimizing \Cref{eq:Y24sec_improved} over the statistical variable $\delta$ (as well as \Cref{eq:Y24sec} over $\delta$ and $\gamma$), we notice, however, that the improvement is even more substantial.

For the upcoming comparisons, we fixed parameters following \Cref{table:common_parameters}.

\begin{table}[ht]
    \centering
    \renewcommand{\arraystretch}{1.5}
    \setlength{\tabcolsep}{8pt}     

    \begin{tabular}{|c|c|}
    \hline
    \textbf{Parameter}
    &
    \textbf{Value} \\
    \hline \hline
    final key bit size
    $\ell$
    &
    $256$ bits
    \\
    \hline
    number of check bits in the parameter testing step
    $k$
    &
    $\frac n2$
    \\
    \hline
    target distinguishing advantage
    &
    $10^{-15}$
    \\
    \hline
    Parameter $\mu$ of the partition check step (when applicable)
    &
    $0.025$ \\
    \hline
    fraction of leaked bits $\vartheta$
    &
    $0.001$
    \\
    \hline
    error correcting code inefficiency\footnotemark
    &
    $1.25$\\
    \hline
    Qubit reception rate
    &
    $80$ kHz\\
    \hline
    \end{tabular}
    \caption{Common fixed parameters for the comparisons}
    \label{table:common_parameters}
\end{table}

\footnotetext{ratio between the considered size $q$ of the syndrome and the optimal size given by the Shannon limit for error reconciliation~\cite{shanon} $h(\delta')\times |I_{\overline{b}}|$ where $h$ denotes the binary entropy function and with $\delta'$ the \textit{actual} error rate, as opposed to the estimated error rate $\delta$.}

\begin{figure}[!ht]
    \centering 
    \includegraphics[width=\textwidth]{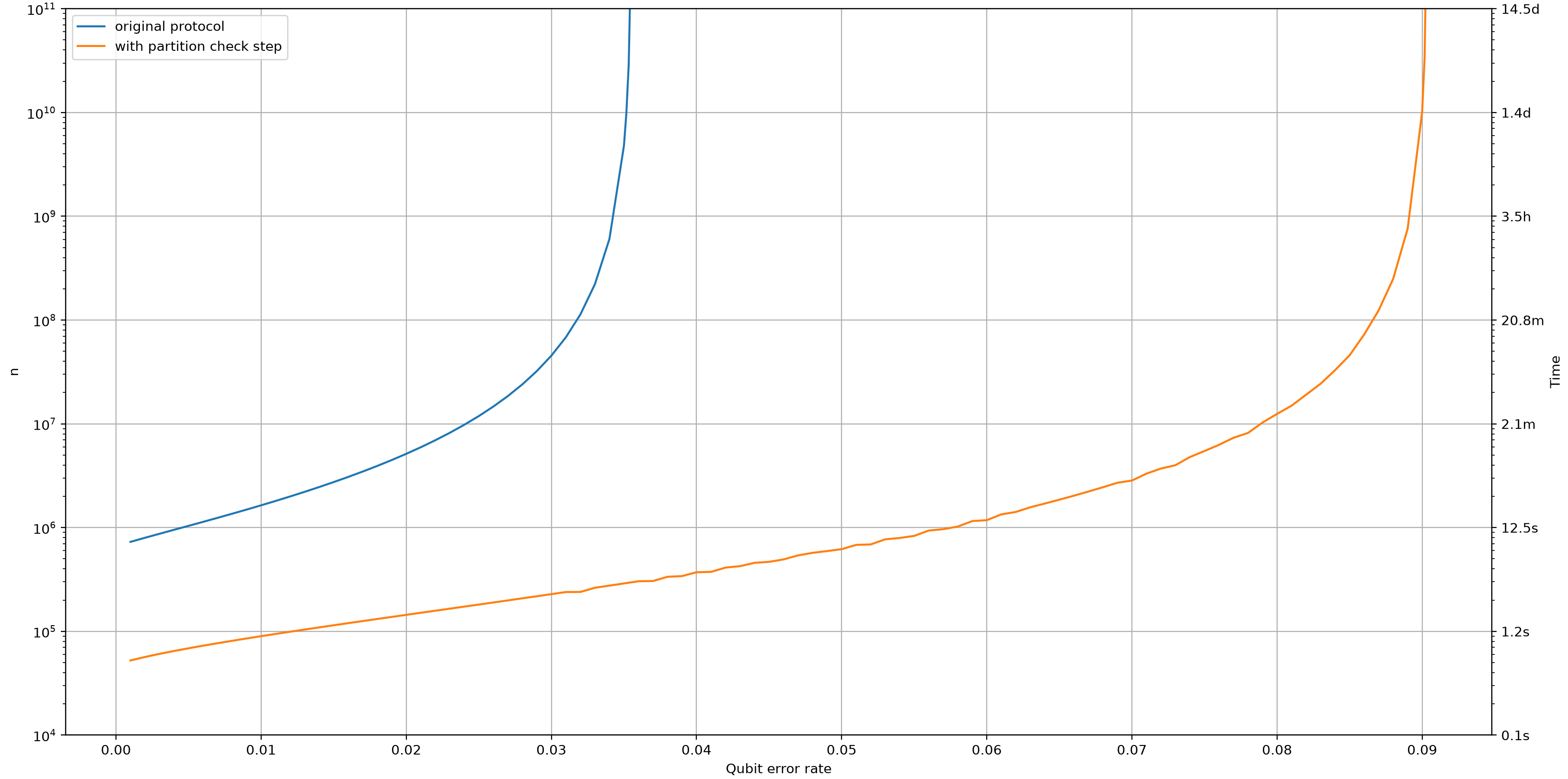}
    \caption{\centering Performance of QOT protocols, without (blue) and with (orange) the partition check step, depending on the qubit error rate.}
    \label{fig:ZK-compare}
\end{figure}

We plot in \Cref{fig:ZK-compare} how the required number $n$ of qubits of the state distribution step varies depending on the qubit error rate, with fixed parameters following \Cref{table:common_parameters}.
The values are minimized over the choice of $\delta$ for the protocol with partition check, and over the choices of $\delta$ and $\gamma$ for the original protocol.
On the right vertical axis, we indicate the corresponding acquisition time with a qubit reception rate of $80$ kHz, as could be reached in the experimental realization of a QOT protocol in \cite{VQ_OT_Tokens}, accounting for losses.
For the version with partition check, the chosen value of $0.025$ for the parameter $\mu$ corresponds to a correctness parameter $\epsilon^\mu$ of about $1.6\times10^{-6}$ for the number of qubits here obtained at $1\%$ error rate:  $89$~$860$.

In particular, the number $n$ of qubits to be received is about $18$ times less with the partition check step than without at $1\%$ qubit error rate, and $199$ times less at $3\%$.
While reducing the quantum channel usage, this also reduces the input size of all the functionalities of the post-processing, some of which exhibit superlinear complexity with respect to their input size. This also reduces the amount of required memory, which has been reported to constitute a potentially blocking factor~\cite{VQ_OT_Tokens}. Finally, in the presence of errors, our proposal drastically increases the maximum tolerable qubit error rate at which the protocol still offers a positive OT rate asymptotically. In our simulations, this threshold appears to lie around $9\%$ with the partition check step, as opposed to about $3.5\%$ without (see \Cref{fig:ZK-compare}).

\section{Estimation and mitigation of the computational overhead}\label{sec:est_and_mitig_overhead}

The zero-knowledge proof underlying our proposal undeniably introduces a significant computational overhead. 
We first provide estimations of this overhead in \Cref{sec:overhead_benchmarks}, showing that it would likely represent the bottleneck for the efficiency of practical implementations. 
Then, we introduce in \Cref{sec:ZK_alternative} an alternative proposal for our partition check step, which reduces the computational work load of the zero-knowledge proof, at the expense of a slight reduction of the transmission gains.

\subsection{Estimation of the computational overhead}\label{sec:overhead_benchmarks}

Current constructions of ZKPs from one-way functions require computations that can be quite extensive, depending on the statement being proven. In the context of quantum oblivious transfer, we are interested in zero-knowledge proof systems that are secure against quantum adversaries, and that only rely on the assumption of one-way function.~\footnote{OT is indeed a cryptographic task that can be achieved classically from structured assumptions, so most of the interest of quantum protocols comes from the possibility of reducing the assumption level to the family of one-way functions.}
Protocols based on the ``MPC-in-the-head'' paradigm~\cite{MPC_in_the_head} thus constitute suitable candidates for an implementation of our proposal.
In addition to relying solely on one-way functions and admitting secure implementations that fall within the scope of Watrous's result for post-quantum zero-knowledge~\cite{Watrous_rewinding}, variants of the ZKBoo protocol~\cite{ZKBoo,KKW18} achieve suitable efficiency. A representative example, which we will use as reference to estimate the computational cost of the partition check step, is the following.
Using the ZKBoo protocol~\cite{ZKBoo}, a zero-knowledge proof of knowledge of a pre-image by SHA-256 with soundness error $\epsilon_S = 2^{-80}$ can be computed, depending on the benchmarks~\cite{ZKBoo,ZKB++}, in $50$ to $70$ ms of sequential, single-core processing of a standard laptop CPU. Such a proof consists in $137$ repetitions of a challenge-response process involving $3$ rounds of communication. The total size of the proof is about $840$ kB. Importantly, for a proof involving the computation of multiple pre-images, the number of repetitions and communication rounds stays the same and the computational cost scales linearly.

In our context, the number of commitments that need to be opened in zero-knowledge is roughly $\frac {n-k}2$.\footnote{Specifically, while an optimized implementation of the proof can avoid opening the commitments of both sets $I_0$ and $I_1$ by only opening the necessary ones in $I_b$, one cannot avoid opening $\vert{}I_b\vert{}$ commitments in zero-knowledge. Furthermore, dummy commitments must be included to avoid leaking the choice bit $b$ from the size of the circuit. This amounts to a total of $\max(\vert{}I_0\vert{}, \vert{}I_1\vert{})$ commitments, which is slightly above $\frac {n-k}2$ commitments for a construction following the structure presented in \Cref{sec:protocol}.}
Consequently, assuming a bit commitment scheme which opening requires one computation of SHA-256, we will consider that computing the zero-knowledge proof of the partition check step represents $\frac {n-k}2$ times $60$ ms of computational work.\\

\Cref{table:ZK_overhead_table} compares the computational overhead of the partition check step, the number of qubits that need to be transmitted in the state distribution step and the corresponding acquisition time in order to run the oblivious transfer protocol with fixed parameters following \Cref{table:common_parameters}. The parameter
$\alpha$ denotes the qubit error rate, $n$ the required number of qubits to be received during the state distribution step (following the notation in the previous sections), $T_{\mathrm{acq}}$ the corresponding acquisition time (assuming a qubit reception rate of $80$ kHz, as stated in \Cref{table:common_parameters}), and $T_{\mathrm{zkp}}$ the computational overhead of the partition check step assuming a linear scaling of $60$ ms per commitment opened in zero-knowledge, according to the sequential ZKBoo benchmarks~\cite{ZKBoo,ZKB++}. The first three columns to the QOT protocol without the particition check step, and the last three columns correspond to the protocol with our proposed partition check step.

\begin{table}[ht]
    \centering
    \renewcommand{\arraystretch}{1.5} 
    \setlength{\tabcolsep}{8pt}       

    \begin{tabular}{|c|c|c|c|c|c|c|}
    \hline
        
      & \multicolumn{3}{|c|}{without the partition check step}
      & \multicolumn{3}{|c|}{with the partition check step}
      \\
    \hline \hline
    $\alpha$  
      & $0.01$  
      & $0.03$
      & $0.06$
      & $0.01$
      & $0.03$
      & $0.06$\\
    \hline
    $n$  
      & $1$ $639$ $026$  
      & $45$ $400$ $070$  
      & $-$
      & $89$ $860$
      & $228$ $350$
      & $1$ $178$ $946$\\
    \hline
    $T_{\mathrm{acq}}$ 
      & $20.5$ s
      & $9.4$ min
      & $-$
      & $1.12$ s  
      & $2.85$ s
      & $14.7$ s  \\
    \hline
    $T_{\mathrm{zkp}}$ 
      & $-$ 
      & $-$
      & $-$
      & $22.5$ min  
      & $57$ min  
      & $295$ min  \\
    \hline
    \end{tabular}

    \vspace{5mm}
    \caption{Performance benchmark of the partition check step}
    \label{table:ZK_overhead_table}
\end{table}

    As shown by \Cref{table:ZK_overhead_table}, without dedicated hardware acceleration for classical computations, the computational overhead of the partition check step would likely exceed the time saved during the state distribution in practical setups, up to the point where the main bottleneck of the full protocol would shift from the state distribution step to the post-processing step.

To account for these situations, we introduce an alternative partition check step in \Cref{sec:ZK_alternative} that significantly reduces the computational overhead, at the expense of moderately increasing the distinguishing advantage.
This alternative acts as a tunable trade-off, allowing the computational overhead of our proposal to be adjusted to accommodate specific hardware capabilities.

It is also worth noting that the computational overhead imposed by the partition check step will naturally be reduced as the efficiency of post-quantum zero-knowledge proofs based on one-way functions improves in the future.

\subsection{Tradeoff in computational overhead} \label{sec:ZK_alternative}

This section introduces an alternative to our proposal designed to reduce the computational overhead of the zero-knowledge proof, at the expense of slightly worsening the security against malicious Bob. This alternative serves as a middle ground between the original QOT protocols and the one with our proposed partition check step. It is particularly well-suited for practical scenarios where the majority of the protocol's execution time is dominated by the zero-knowledge computation rather than the quantum channel usage. This alternative is parametrized by a parameter $k'$, representing the fixed number of commitments opened in the zero-knowledge proof. Consequently, by tuning $k'$, one can precisely adjust the computational overhead to align with specific operational requirements.

We first define the proposed alternative in \Cref{sec:alternative_def}, before demonstrating how it modifies the distinguishing advantage against a malicious Bob in \Cref{sec:alternative_sec}, and comparing its performances with the original protocol and the initial proposal in \Cref{sec:alternative_perfs}.\\ 

\subsubsection{Alternative proposal}\label{sec:alternative_def}\

The underlying concept mimics the cut-and-choose strategy used in the original QOT protocol, where Alice verifies Bob's honesty on most measurements by checking only a randomly selected subset, leaving the rest for the remainder of the protocol. 
In our context, Bob proves a statement similar to \Cref{ZK-statement} in zero-knowledge, but restricted to a subset of $k'$ bits randomly chosen by Alice. 
This allows the zero-knowledge proof to involve significantly fewer commitment openings, thereby alleviating the corresponding computational overhead. However, this approach inherently allows Bob to \emph{slightly} diverge from the honest set partitioning step while maintaining a low probability of getting caught. Concretely, compared to the initially proposed partition check step, this alternative proposal provides a moderately looser bound on a malicious Bob's entropy on the secret, according to a statistical deviation that accounts for such an attack.
We refer to this variant as the \emph{partial} partition check step.

\begin{enumerate}[label=4\alph*, start=2]
    \item \textbf{Partial partition check:}\\
    \vspace{3mm}\hspace{4mm}\textit{Parameters: $\mu \in [0, 1]$; $k'\in \llbracket 1, \frac{n-k}{2}(1-\mu)\rrbracket$}\\
    If $|I_0| \le \frac{n-k}{2}(1-\mu)$ or $|I_1| \le \frac{n-k}{2}(1-\mu)$, the protocol aborts.

    Alice randomly samples and announces two challenge subsets $J_0$ and $J_1$ of $I_0$ and $I_1$ respectively, each of size $k'$.
    Bob checks that $J_0\subset I_0$ and that $J_1\subset I_1$ (aborting if not), before computing and sending to Alice a zero-knowledge proof $\pi$ that 
    \begin{equation} \label{ZK-statement_alt}
    \exists b,(o_i)_{i\in J_{\bar b}} \forall i \in J_{\bar b}, \textbf{Open(}c_i^\theta,o_i\textbf{)}=\overline{\theta_i^A}
    \end{equation}
    
    Alice verifies the proof, aborting if it fails.
\end{enumerate}

\subsubsection{Security against malicious Bob}\label{sec:alternative_sec}

The following \Cref{thm:Y24_alternative} shows the security bound against a malicious Bob for a QOT protocol with the partial partition check step. 

\begin{theorem} \label{thm:Y24_alternative}
    Consider a QOT protocol following the structure described in \Cref{sec:protocol} with statistically binding commitments and the partial partition check step parametrized by $\mu$ and $k'$.
    Then, for any $\delta \in [0, \frac12]$ and any $\nu\in[0,1]$, the protocol is $\epsilon_A'$-secure against a malicious Bob in the sense of \Cref{def:OT_sec} with
    \begin{equation} \label{eq:Y24sec_alternative}
        \epsilon_A' \leq
        \frac{1}{2} \cdot 
        2^{-\frac{1}{2}\big[\frac{n-k}{2} \big( (1-\mu - 2\vartheta) \times ( 1 - h(\delta+\alpha) ) - \nu(1-\mu) \big)
        -\ell-q\big]}
        + \sqrt6 e^{-\frac{\delta^2k}{100}}
        + (1-\nu)^{k'}
        + \epsilon_S
    \end{equation}
    where:
    \begin{itemize}
        \item $\alpha$ is the qubit error rate,
        \item $q$ is the size of the syndromes used for the error-correcting step,
        \item $\vartheta$ is the proportion of leaked bits (for instance due to plausible multi-photon events).
        \item $\epsilon_S$ is the soundness error of the zero-knowledge proof
    \end{itemize}
\end{theorem}

\begin{proof}[Proof sketch.]
We use the same notations as for the proof of \Cref{thm:malicious_bob_Y24_improved} and once again reproduce a similar proof as in~\cite{BF12}. We consider the equivalent entanglement-based protocol $QOT^*$.

Assume that Bob dishonestly performed the set partitioning step and included in \emph{both} $I_{0}$ and $I_1$ a proportion strictly greater than $\nu>0$ of indices $i$ such that $\textbf{Open(}c_i^\theta,o_i\textbf{)}\ne\overline{\theta_i^A}$. For all $b\in\{0, 1\}$, we define $K_{\bar b} = |I_{\bar b}| \times (1-\nu)$ the upper bound on the number of indices $i$ among $I_{\bar b}$ such that $\textbf{Open(}c_i^\theta,o_i\textbf{)}=\overline{\theta_i^A}$.
This assumption yields, for all $b\in\{0, 1\}$,
\begin{equation} \label{eq:dist_assumption_alt_noiseless}
    d(\theta^A_{I_{\bar b}}, \theta^B_{I_{\bar b}}) < K_{\bar b}.
\end{equation}
Given this assumption, for all $b\in\{0, 1\}$, the probability over the random choices of $J_0$ and $J_1$ that Statement~\ref{ZK-statement_alt} holds is at most \footnote{This holds as long as $K_{\bar b}>k'$. If not, Statement~\ref{ZK-statement_alt} simply does not hold, regardless of the choices of $J_0$ and $J_1$.}: 
\begin{equation}
    \frac{K_{\bar b}}{|I_{\bar b}|} \times \frac{K_{\bar b}-1}{|I_{\bar b}|-1} \times \dots \times \frac{K_{\bar b}-(k'-1)}{|I_{\bar b}|-(k'-1)} 
    \le \left(\frac{K_{\bar b}}{|I_{\bar b}|}\right)^{k'} 
    = (1-\nu)^{k'}.
\end{equation}

Adding this probability to the soundness error of the zero-knowledge scheme, we can bound the probability that the protocol proceeds despite \Cref{eq:dist_assumption_alt_noiseless} being verified by $(1-\nu)^{k'}+\epsilon_S$.

We book-keep this probability and assume for the remainder of the proof that there exists a bit $b\in\{0, 1\}$ such that
\begin{equation} \label{eq:wt_theta_I1-c_leaks_improved_alt}
    d(\theta^A_{I_{\bar b}}, \theta^B_{I_{\bar b}}) \ge |I_{\bar b}| \times (1-\nu).
\end{equation}

We assume external leakage of a proportion $\vartheta$ of Alice's secrets among the $n-k$ indices of $I_0\cup I_1$. Considering that malicious Bob is taking maximum advantage of these $\vartheta(n-k)$ leaks by putting all of them in $I_{\bar b}$, we will focus on the $|I_{\bar b}|-\vartheta(n-k)$ bits of $I_{\bar b}$ that did not leak. We call $S$ this subset.

\Cref{eq:wt_theta_I1-c_leaks_improved_alt} implies that
\begin{align}
    d(\theta^A_{S}, \theta^B_{S}) 
    &\ge |S| - \nu|I_{\bar b}|.
\end{align}

Let $X_S$ be the random variable obtained by measuring the qubits of indices in $S$ in basis $H^{{\theta_S}} \{\ket0, \ket1\}^{\otimes|S|}$.
Due to the structure of the parameter testing step, \Cref{thm:BF12_Hmin_bound_imported} applies to $X_S$, yielding
\begin{align}
    H_{min} (X_{S}|E) &\ge  d(\theta^A_{S}, \theta^B_{S}) - h(\delta+\alpha) \times |S|\\
    &\ge  |S|\times(1- h(\delta+\alpha)) - \nu|I_{\bar b}| \\
    &\ge  (|I_{\bar b}|-\vartheta(n-k))\times(1- h(\delta+\alpha)) - \nu|I_{\bar b}|.
\end{align}
Since $\frac{n-k}{2}(1-\mu) \le |I_{\bar b}|$,\footnote{If $\nu \le (1-h(\delta+\alpha))$, the implication is direct. If not, the bound is always true since min-entropies are positive.}
\begin{align}
    H_{min} (X_{S}|E) &\ge \frac{n-k}{2}\big[(1-\mu-2\vartheta)\times(1- h(\delta+\alpha)) - \nu(1-\mu)\big].
\end{align}
Finally, because $S$ is a subset of $I_{\bar b}$, $H_{min} (X_{\bar b}|E) \ge H_{min} (X_{S}|E)$.

\Cref{eq:Y24sec_alternative} follows from the leftover hashing lemma (\Cref{lem:leftover_hashing}), considering the leakage $q$ from the syndrome of the error correction and the error probabilities accumulated along the way.

\end{proof}

As for \Cref{thm:malicious_bob_Y24_improved}, setting the parameters $\alpha$, $\vartheta$ and $q$ to $0$ gives a noiseless version of \Cref{thm:Y24_alternative}.

\begin{corollary}
    In a perfect setting without losses nor errors, consider a QOT protocol following the structure described in \Cref{sec:protocol} with statistically binding commitments and the partial partition check step parametrized by $\mu$ and $k'$.
    Then, for any $\delta \in [0, \frac12]$ and any $\nu\in[0,1]$, the protocol is $\epsilon_{A, \text{noiseless}}'$-secure against a malicious Bob in the sense of \Cref{def:OT_sec} with
    \begin{equation} 
        \epsilon_{A, \text{noiseless}}' \leq
        \frac{1}{2} \cdot 
        2^{-\frac{1}{2}\big[\frac{n-k}{2} \big( (1-\mu) \times ( 1 - h(\delta) ) - \nu(1-\mu) \big)
        -\ell\big]}
        + \sqrt6 e^{-\frac{\delta^2k}{100}}
        + (1-\nu)^{k'}
        + \epsilon_S
    \end{equation}
    where:
    \begin{itemize}
        \item $q$ is the size of the syndromes used for the error-correcting step,
        \item $\epsilon_S$ is the soundness error of the zero-knowledge proof
    \end{itemize}
\end{corollary}

\subsubsection{Performance}\label{sec:alternative_perfs}

\begin{figure}[!ht]
    \centering 
    \includegraphics[width=\textwidth]{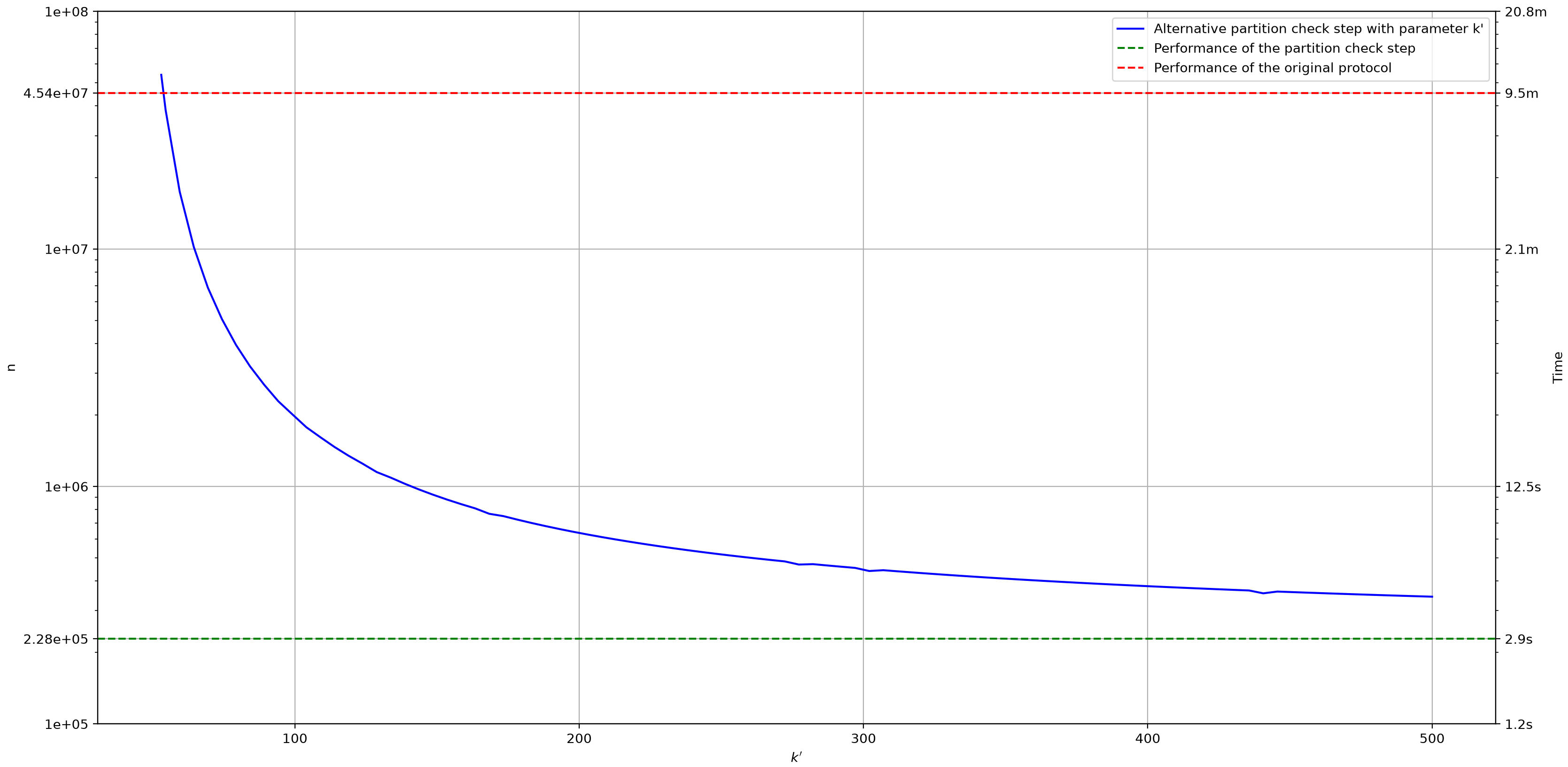}
    \caption{\centering Performance of the partial partition check step.}
    \label{fig:ZK_alternative_perfs}
\end{figure}

\Cref{fig:ZK_alternative_perfs} shows how the number of qubits required in the state distribution step behaves depending on the parameter $k'$ of the partition check step, with fixed parameters following \Cref{table:common_parameters} and a qubit error rate $\alpha = 0.03$.
The values are minimized over the choice of $\delta$ and $\nu$.
On the right vertical axis is shown the corresponding acquisition time (with a reception rate of $80$ kHz as stated in \Cref{table:common_parameters}).
The dashed horizontal lines represent the same number $n$ for the protocol without the partition check step (red line), and with the (full) partition check step (green line). These can be thought as upper and lower bounds, respectively.

We note that while the performance of the alternative partition check step converges (close) to the one of the initial partition check step as $k'$ grows, the partition check step with small $k'$ underperforms the original protocol for small enough values of $k'$ (for $k'<54$ with these parameters). This is explained by the fact that with such a low number of commitments opened in zero-knowledge, it cannot even be guaranteed, with good enough confidence, that Bob honestly split half his knowledge among the two sets $I_0$ and $I_1$ (The term $(1-\nu)^k$ indeed reaches $2^{-50} \approx 8.8\times10^{-16}$ with $\nu=\frac12$ and $k'=50$, which already accounts for almost all the security budget here imposed of $\epsilon < 10^{-15}$). On the other hand, that very statement of ``half honesty'' can be obtained for free from a pigeonhole principle as used in the security proof of \cite{BF12}.

The values are plotted until $k'=500$ which corresponds to a computation time of about $30$s for the partial partition check step according to the benchmarks of the ZKBoo protocol \cite{ZKBoo,ZKB++}.

\medskip

The following \Cref{table:ZK_alt_overhead_table} takes the previous performance benchmark shown in \Cref{table:ZK_overhead_table}, with the same parameters, but replaces the partition check step with the partial alternative, parametrized by a parameter $k'$ carefully chosen so that the computational overhead of the (partial) partition check exactly matches the acquisition time of the state distribution.

\begin{table}[ht]
    \centering
    \renewcommand{\arraystretch}{1.5} 
    \setlength{\tabcolsep}{8pt}        

    \begin{tabular}{|c|c|c|c|c|c|c|}
    \hline
        
      & \multicolumn{3}{|c|}{without the partition check step}
      & \multicolumn{3}{|c|}{with the partial partition check step}
      \\
    \hline \hline
    $\alpha$  
      & $0.01$  
      & $0.03$
      & $0.06$
      & $0.01$
      & $0.03$
      & $0.06$\\
    \hline
    $n$  
      & $1$ $639$ $026$  
      & $45$ $400$ $070$  
      & $-$
      & $465$ $600$
      & $796$ $800$
      & $2$ $561$ $082$\\
    \hline
    $T_{\mathrm{acq}}$ 
      & $20.5$ s
      & $9.4$ min
      & $-$
      & $5.8$ s  
      & $10.0$ s
      & $32.0$ s  \\
    \hline
    $T_{\mathrm{zkp}}$ 
      & $-$ 
      & $-$
      & $-$
      & $5.8$ s
      & $10.0$ s
      & $32.0$ s  \\
    \hline
    \end{tabular}

    \vspace{5mm}
    \caption{Performance benchmark of the partial partition check step}
    \label{table:ZK_alt_overhead_table}
\end{table}

We observe an overall improvement in the performance, especially for higher error regimes.

\section{Future directions} \label{sec:future_directions}

Building upon the enhancements proposed in this work, several promising avenues remain for further improving QOT protocols.

In recent constructions~\cite{OTfromOWF1,OTfromOWF2,Y24}, composable QOT from OWF in the plain model is achieved using \textit{equivocal} and \textit{extractable} bit commitments. These properties together enable security under a stronger, simulation-based security definition which provides composability~\cite{DFL09}.
However, building equivocal and extractable bit commitments requires extensive quantum communication, causing a big overhead in such composable settings. 

\medskip

A first interesting direction to investigate is the use of zero-knowledge proofs in the parameter testing step in these commitment sub-protocols. In a context where ZKPs are already integrated in the protocol, the additional implementation cost would be amortized. 
ZKPs could indeed be used to obtain an equivocal commitment from a single use of a standard (statistically binding) commitment by changing the opening process of a commitment scheme into a zero-knowledge proof that the commitment can be opened onto a claimed value.
Such a construction could turn out to be of significant interest considering that other constructions for equivocal commitments, used in ~\cite{OTfromOWF1,OTfromOWF2,Y24}, require repeating the commitment process sequentially multiple times.\\

Another promising direction is to explore the possibility of introducing 
proofs of knowledge as an alternative for extractable bit commitments to obtain composability. 
In essence, a zero-knowledge proof of knowledge (ZKPoK) is a stronger variant of a standard ZKP, guaranteeing that a prover who successfully convinces a verifier must explicitly possess the underlying witness. In our proposed partition check step, this witness incorporates the receiver's secret choice $b$. Furthermore, the formal definition of a proof of knowledge relies on extraction and simulation paradigms analogous to the composable security definitions of oblivious transfer. By leveraging this conceptual alignment, it may be possible to circumvent the need for extractable commitments entirely, achieving composability relying solely on ZKPoKs and equivocal commitments. Such a result would yield substantial efficiency gains, given the significant quantum communication overhead imposed by existing extractable commitment schemes.
However, this approach faces two major interconnected hurdles. 
First, whether one-way functions imply, in a black-box manner, proofs of knowledge secure against quantum adversaries remains, to the best of our knowledge, an open question. Secure constructions of such a primitive have been proposed from seemingly close assumptions, like quantum-secure 1-1 one-way functions~\cite{Unruh_ZKPoK}, collapsing hash functions or super-polynomially secure one-way functions~\cite{ZKPoK_revisited}. 
Second, the witness extraction enabled by proofs of knowledge is not the sole requirement in order to satisfy composable security against a malicious receiver. The receiver's quantum state must also remain unperturbed after extraction, a condition that is not guaranteed by standard proof of knowledge constructions. The notion of undetectable extraction has been introduced recently by Lombardi, Ma and Spooner~\cite{ZKPoK_revisited} in a different context, but extensive work is still required to study the applicability of this concept to the framework of oblivious transfer.
Conversely, if these hurdles cannot be overcome, establishing a formal impossibility result would clarify the minimal assumptions required to achieve composable security in this setting.\\

\newpage
\bibliographystyle{plain}
\bibliography{MyRef}

\end{document}